\documentclass[twocolumn,aps,prl,reprint, superscriptaddress,longbibliography]{revtex4-2}

\usepackage{amsmath}
\usepackage{amssymb}
\usepackage{graphicx}
\usepackage{color}
\usepackage{hyperref}
\usepackage{xcolor}
\hypersetup{
    colorlinks,
    linkcolor={blue!80!black},
    citecolor={blue!80!black},
    urlcolor={blue!80!black}
}
\usepackage[caption=false]{subfig}
\usepackage[colorinlistoftodos]{todonotes}
\usepackage{lipsum}

\usepackage{tabularx}
\usepackage{booktabs}
\usepackage{multirow}
\usepackage{enumitem}

\usepackage{lipsum}
\usepackage{todonotes}
\usepackage{titlesec}

\graphicspath{ {main_figures/} {supplement_figures/} }



\begin{document}
\titleformat{\section}{\bfseries\small\centering}{\thesection.}{1em}{\MakeUppercase}

\newcommand{\Exa}{\frac{\partial \alpha_\mathrm{res}^\mathrm{a}(\mathbf{r}_i)}{\partial x} }
\newcommand{\Exb}{\frac{\partial \alpha_\mathrm{res}^\mathrm{b}(\mathbf{r}_i)}{\partial x} }
\newcommand{\Eya}{\frac{\partial \alpha_\mathrm{res}^\mathrm{a}(\mathbf{r}_i)}{\partial y} }
\newcommand{\Eyb}{\frac{\partial \alpha_\mathrm{res}^\mathrm{b}(\mathbf{r}_i)}{\partial y} }

\title{High-impedance resonators for strong coupling to an electron on helium}

\begin{abstract}
    The in-plane motion of an electron on helium can couple to superconducting microwave resonators via electrical dipole coupling, offering a robust and rapid readout scheme. In previous efforts, microwave resonator designs for electrons on helium have lacked the coupling strength to reach the strong coupling regime, where coherent quantum effects outlast both electron and resonator decoherence rates. High-impedance superconducting microwave resonators offer a path to strong coupling, but integrating such resonators with electrons on helium remains an outstanding challenge. Here, we introduce a high-impedance resonator design compatible with strong coupling to electrons on helium. We fabricate and measure titanium nitride resonators with median internal quality factors of $3.9\times 10^5$ and average impedance of 2.5~k$\Omega$, promising a seven-fold increase in coupling strength compared with standard 50~$\Omega$ resonators. Additionally, we develop a simplified resonator model from the capacitance matrix and sheet inductance that accurately predicts the mode frequencies, significantly simplifying the design process of future resonators for investigating quantum effects with electrons on helium.
\end{abstract}

\author{G.~Koolstra}
\email{koolstragerwin@gmail.com}
\affiliation{EeroQ Corporation, Chicago, Illinois, 60651, USA}
\author{E.~O.~Glen}
\affiliation{EeroQ Corporation, Chicago, Illinois, 60651, USA}
\author{N.~R.~Beysengulov}
\affiliation{EeroQ Corporation, Chicago, Illinois, 60651, USA}
\author{H.~Byeon}
\affiliation{EeroQ Corporation, Chicago, Illinois, 60651, USA}
\author{K.~E.~Castoria}
\affiliation{EeroQ Corporation, Chicago, Illinois, 60651, USA}
\author{M.~Sammon}
\affiliation{EeroQ Corporation, Chicago, Illinois, 60651, USA}
\author{B.~Dizdar}
\affiliation{James Franck Institute and Department of Physics, University of Chicago, Chicago, Illinois 60637, USA}
\author{C.~S.~Wang}
\affiliation{James Franck Institute and Department of Physics, University of Chicago, Chicago, Illinois 60637, USA}
\author{D.~I.~Schuster}
\affiliation{James Franck Institute and Department of Physics, University of Chicago, Chicago, Illinois 60637, USA}
\affiliation{Department of Applied Physics, Stanford University, Stanford, California 94305, USA}
\author{S.~A.~Lyon}
\affiliation{EeroQ Corporation, Chicago, Illinois, 60651, USA}
\author{J.~Pollanen}
\affiliation{EeroQ Corporation, Chicago, Illinois, 60651, USA}
\author{D.~G.~Rees}
\affiliation{EeroQ Corporation, Chicago, Illinois, 60651, USA}

\date{\today}

\maketitle

\section{Introduction}
Electrons on helium are a unique two-dimensional system with potential for quantum computing due to their ability to scale towards large quantum processors \cite{Platzman1999, Dykman2003, Lyon2006, Schuster2010, Kawakami2023, Jennings2024}. Developing quantum computers with electrons on helium first requires an approach for fast, high-fidelity readout. Among the various readout techniques \cite{Kawakami2019, Papageorgiou2005, Bradbury2011}, coupling the in-plane motion of electrons on helium to microwave resonators \cite{Schuster2010} is particularly promising, because the electrical dipole coupling strength allows for submicrosecond readout via circuit quantum electrodynamics (cQED) techniques \cite{Blais2021, Clerk2020}.

However, high-fidelity readout using the cQED framework requires strong coupling between a microwave photon and the in-plane motional state, which means that the coherent interaction between resonator photons and the in-plane electron motion $g$ occurs on a faster timescale than microwave photon loss $\kappa$, or electron decoherence $\Gamma$ \cite{Wallraff2004, Zhou2022, Zhou2024, Clerk2020}. While proposals for strong coupling to electrons on helium were developed over a decade ago \cite{Schuster2010}, current state-of-the art experiments have achieved a coupling strength of $g/2\pi \approx 5$~MHz and electron damping rate of $\Gamma/2\pi \approx 77$~MHz, still deep in the weak coupling regime \cite{Koolstra2019}. Therefore, the coupling strength must be increased by at least an order of magnitude to reach the strong coupling regime. 

High-impedance resonators can bridge the gap between weak and strong coupling, because the resonator’s vacuum voltage fluctuations scale as $\sqrt{Z_\mathrm{res}}$ \cite{Clerk2020}. Specifically, the coupling is
\begin{equation}
    g = \frac{1}{2} e \mathcal{E}_x \omega_0 \sqrt{\frac{Z_\mathrm{res}}{m_e \omega_e}}, \label{eq:coupling_strength}
\end{equation}
where $e$ and $m_e$ are the electron charge and mass, $\mathcal{E}_x$ is the microwave electric field per volt at the location of the electron, $\omega_0$ and $\omega_e$ are the frequency of the resonator and in-plane electron motion, and $Z_\mathrm{res}$ is the resonator impedance \cite{Koolstra2019}. Thus, strong coupling to single electrons can be engineered by increasing the resonator impedance $Z_\mathrm{res}$, provided the internal photon loss rate $\kappa_i/2\pi < 1$~MHz remains small, which can be challenging in the presence of the gate electrodes needed to trap electrons \cite{HarveyCollard2020}. High-impedance resonators have been realized using Josephson junction arrays \cite{MaslukPRL2012, Stockklauser2017} or granular aluminum \cite{MaleevaNatComm2018, Grunhaupt2018}. Alternatively, resonators made from high-kinetic inductance superconductors such as titanium nitride (TiN) are attractive due to ease of fabrication and low-loss at microwave frequencies \cite{Shearrow2018}. 

While high-kinetic inductance resonators have been used for strong coupling in semiconductors and other hybrid quantum systems \cite{Stockklauser2017, Samkharadze2018, Yu2023}, these resonator designs do not readily comply with the experimental requirements for electrons on helium. In addition, previous low-impedance resonator designs tailored to electrons on helium are not optimal, because they simultaneously coupled to large electron ensembles \cite{Yang2016, Koolstra2019}, which was identified as a possible source of increased $\Gamma$ \cite{KoolstraThesis}. Therefore, a new resonator design is necessary, which requires (i) an integrated dot for electron trapping where the resonator's electric field $\mathcal{E}_x$ is strong, while the coupling to stray electrons outside the dot is minimized, and (ii) a bias-able resonator center pin electrode to control the motional frequency $\omega_e$ via the trapping potential. 

Here, we introduce and experimentally demonstrate a realization of a high-impedance resonator circuit tailored to electrons on helium. Our symmetrically coupled resonator (SCR) design, which is the analog of an inductively coupled resonator design used to increase magnetic dipole coupling \cite{Eichler2017, McKenzie-Sell2019}, complies with the above requirements for strong coupling to electrons on helium. To show this, we fabricate and measure TiN resonators with average extracted impedances of approximately 2.5~k$\Omega$ and average loss rates of $\kappa_i/2\pi = 11.7$~kHz, promising a sevenfold increase in $g$. Furthermore, we develop a theoretical description for the SCR coupling to $n$ electron clusters, which shows that SCRs can be used as sensitive electron detectors, with $g$ exceeding tens of megahertz. These results show that TiN SCRs bring strong coupling to electrons on helium within reach.

This article is organized as follows. In Sec.~\ref{sec:theoretical_description} we analyze the design of symmetrically coupled resonators using the Lagrangian formalism. We simulate the eigenmodes and demonstrate that these resonators have the desired sensitivity for coupling to few electrons on helium. Next, in Sec.~\ref{sec:experimental_realization}, we demonstrate our implementation of symmetrically coupled resonators. We characterize the microwave resonators in a dilution refrigerator and using a theoretical model, we infer the resonator characteristic impedance $Z_\mathrm{res}$. Finally, in Sec.~\ref{sec:discussion} and \ref{sec:conclusion} we summarize the prospects of using symmetrically coupled resonators for strong coupling to an electron on helium. 

\section{Theoretical description} \label{sec:theoretical_description}

\begin{figure}[tbh]
    \centering
    \includegraphics[width=\columnwidth]{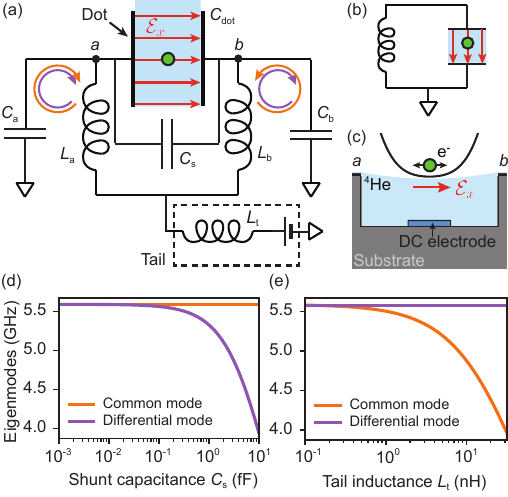}
    \caption{Resonator design (a) Circuit schematic for a symmetrically coupled resonator (SCR) that couples sensitively to an electrons on helium. The resonator eigenmodes are clockwise and counterclockwise currents in the left and right halves of the circuit. The differential mode (purple) generates a microwave electric field $\mathcal{E}_x$ in the dot region, which can couple to trapped electrons (green dots) on helium (blue). The SCR design also allows for applying a dc bias to the resonator center pin, with minimal consequences to the differential mode. (b) Conventional single-ended coupling scheme where one of the capacitor plates remains grounded.  (c) Cross sectional schematic of an electron confined by the trapping potential (black curve) generated by voltages on any submerged dc electrodes or resonator center pin. Nodes $a$ and $b$ refer to the same nodes shown in (a). (d) Increasing $C_x$ lifts the degeneracy between common and differential modes, but also decreases the differential mode impedance. (e) Alternatively, adding an inductor $L_t$ to the resonator tail suppresses the common mode frequency of this circuit, without compromising the differential mode impedance.
}
    \label{fig:fig1}
\end{figure}

The SCR design shown in Fig.~\ref{fig:fig1}a consists of two parallel $LC$ resonators coupled via a capacitance $C_\mathrm{dot}$. The space between the plates of $C_\mathrm{dot}$ is the only region where electrons are intended to be trapped and interact with the microwave electric field of the coupled modes $\mathcal{E}_{x}$. This coupled design has two advantages over a single-ended coupling scheme (Fig.~\ref{fig:fig1}b and e.g. Ref.~\cite{Yang2016}). First, the coupled eigenmodes give rise to a larger voltage drop across $C_\mathrm{dot}$, which boosts the electron-photon coupling $g$ by a factor of $\sqrt{2}$ compared with a single-ended coupling scheme (see App.~\ref{app:coupling_strength_derivation}). Secondly, we exploit the symmetry of the coupled eigenmodes to introduce a dc biasing structure (``Tail'' in Fig.~\ref{fig:fig1}a) that does not affect the main microwave mode. In future devices, this tail can be biased to control the electrostatic potential for trapped electrons in the dot (Fig.~\ref{fig:fig1}c). 

The coupled SCR eigenmodes are co-rotating (Fig.~\ref{fig:fig1}a, purple) and counter-rotating (Fig.~\ref{fig:fig1}a, orange) currents, referred to as the differential and common mode, respectively. These modes originate from capacitive hybridization of the left and right $LC$ resonators, which follows from the circuit Lagrangian (ignoring the tail for simplicity) \cite{Vool2017}:
\begin{align}
    \mathcal{L}_\mathrm{res} &= \frac{C_a}{2} \dot{\phi}_a^2 + \frac{C_b}{2} \dot{\phi}_{b}^2 + \frac{C_x (\dot{\phi}_a - \dot{\phi}_b) ^2}{2} - \frac{\phi_a^2}{2L_a} - \frac{\phi_b^2}{2L_b},
\end{align}
where $\phi_{a,b}$ are the node fluxes, $C_x = C_s + C_\mathrm{dot}$ is the cross capacitance between nodes $a$ and $b$, and $C_s$ is an additional shunt capacitance. In the limit of a fully symmetric circuit ($L_a = L_b = L$, $C_a = C_b = C$), the voltage across the dot oscillates either in-phase $\dot{\phi}_c = (\dot{\phi}_a + \dot{\phi}_b)/\sqrt{2}$ for the common mode, or out-of-phase $\dot{\phi}_d = (\dot{\phi}_a - \dot{\phi}_b)/\sqrt{2}$ for the differential mode. The corresponding eigenfrequencies for the common and differential mode are
\begin{align}
    &\omega_{0, c} = 1/\sqrt{LC} \\
    &\omega_{0, d} = 1/\sqrt{L(C + 2 C_x)}.
\end{align}
Since only the differential mode produces a microwave field $\mathcal{E}_x$ that allows for coupling to a single electron on helium, it is the main coupled mode of interest. 

To use the differential mode for coupling to electrons, it must be distinguishable from the common mode. Fig.~\ref{fig:fig1}d shows that for small $C_x$ both the common and differential modes are degenerate, which makes independently addressing either mode experimentally challenging. Although the degeneracy can be lifted by choosing a design with larger $C_x$, this also decreases the differential mode characteristic impedance. Therefore, increasing $C_x$ to lift the degeneracy is not preferable.

An alternative method to lift the degeneracy involves impedance engineering of the resonator tail. Fig.~\ref{fig:fig1}e shows that increasing the tail inductance $L_t$ lifts the degeneracy by decreasing the common mode frequency, while the differential mode remains unaffected. More precisely, in App.~\ref{app:inductive_tail} we show that for a fully symmetric circuit the new eigenfrequencies are given by 
\begin{align}
    &\omega_{0, c} = 1/ \sqrt{C (L + 2 L_t)} \\ 
    &\omega_{0, d} = 1/\sqrt{L (C + 2 C_x)}.
\end{align}
In addition, the differential mode impedance is independent of $L_t$:
\begin{align}
    Z_{\mathrm{res},d} = \sqrt{\frac{L}{C + 2 C_x}}.
    \label{eq:diff_impedance}
\end{align}
Therefore, with tail-engineering we achieve a sufficient mode splitting without sacrificing the differential mode impedance.

\begin{figure}[t]
    \centering
    \includegraphics[width=\columnwidth]{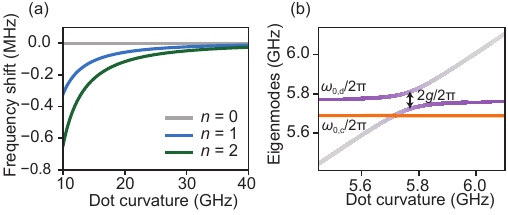}
    \caption{Simulating electron coupling to the circuit in Fig~\ref{fig:fig1}a with parameters from Table~\ref{tab:q3d_simulation}. Electrons are confined in the dot by a one-dimensional harmonic electrostatic potential with curvature $\omega_\mathrm{dot}/2\pi$. For a single electron the dot curvature is equal to the motional frequency. (a) Counting electrons with the differential mode in the dispersive regime. Distinct resonator signatures appear for $n=0, 1, 2$ electrons, showing this resonator design is capable of sensing few electrons off-resonantly. In the case of a harmonic dot, the dispersive shift for $n = 2$ electrons is exactly twice the single electron shift. (b) For $n=1$ electron near resonance, an avoided level-crossing for the in-plane motion with the differential mode (purple) shows potential for strong coupling, while the common mode (orange) remains uncoupled as expected.}
    \label{fig:fig2}
\end{figure}

When electrons on helium are trapped in the dot, the resonator acts as a sensitive electron detector. A resonance frequency shift indicates the presence of electrons, and the magnitude of the shift depends on the number of electrons, curvature of the electrostatic potential and the coupling strength. To predict the performance of the resonator as an electron detector, we extend the theoretical framework developed in Ref.~\cite{Yang2016} to include the electron coupling to both plates of the capacitor $C_\mathrm{dot}$ (see App.~~\ref{app:hamiltonian_formalism_coupling}). 

Using this extended framework, we numerically diagonalize the total Hamiltonian, which includes the resonator with inductive tail, electron motion and coupling Hamiltonians, for two experimentally relevant scenarios: counting electrons during (un)loading of the dot, and demonstrating strong coupling. We assume realistic circuit parameters shown in Appendix~\ref{app:exp_setup} and Table~\ref{tab:q3d_simulation}, ignore electron decoherence, and assume that the electrons are constrained in a one-dimensional harmonic electrostatic potential 
\begin{equation}
    -e V_\mathrm{dc} = \frac{1}{2} m_e \omega_\mathrm{dot}^2 x^2,
\end{equation}
where $\omega_\mathrm{dot}/ 2 \pi$ is the dot curvature that defines the electron in-plane motional frequency. The dot curvature can be controlled in-situ with dc voltages. Even though future trapping potentials will be designed to have anharmonicity, the linear classical treatment we present here is sufficient to demonstrate the key features for coupling to electrons on helium. In addition, the theory detailed in App.~\ref{app:hamiltonian_formalism_coupling} can easily be adjusted to future resonator designs with more than two circuit nodes, or transmission line resonators. 

As electrons enter or leave the dot, the dot curvature is often far detuned from the resonator frequency $\omega_{0,d}$, reducing the resonator's sensitivity for counting electrons \cite{Koolstra2019}. Fig.~\ref{fig:fig2}a shows that even if the in-plane electron motional frequency is detuned by several gigahertz, the expected frequency shifts for $n=1,2$ electrons are approximately 0.1~MHz. Since linewidths of superconducting resonators are typically $<1$~MHz, such shifts constitute a significant fraction of the linewidth. Therefore, the resolution predicted by Fig.~\ref{fig:fig2}a is more than sufficient for counting few electrons in the dot.

Since experimentally demonstrating strong coupling requires bringing the electron into resonance with the resonator, we also study the simulated resonant response for a single electron. The results in Fig.~\ref{fig:fig2}b show that the differential resonator mode and electron motion undergo an avoided level crossing when $\omega_\mathrm{dot} = \omega_{0,d}$, while the common mode remains unaffected. This confirms that only the differential mode is suitable for coupling to electrons on helium. In addition, the size of the avoided crossing is $2g$ and for a realistic value of $\mathcal{E}_x = 0.25$~$\mu$m$^{-1}$ the coupling strength is $g/2\pi = 43$~MHz. The simulated results in the dispersive and resonant regime show that SCR resonators theoretically produce the required features for strong coupling.

\section{Experimental realization of resonator} \label{sec:experimental_realization}
Given that we have developed a thorough theoretical understanding of the SCR design and coupling model, we next test the experimental feasibility of these devices and demonstrate excellent agreement of the eigenmode properties with our theoretical model.

\subsection{Resonator design and fabrication}
\begin{figure}[htbp]
    \centering
    \includegraphics[width=\columnwidth]{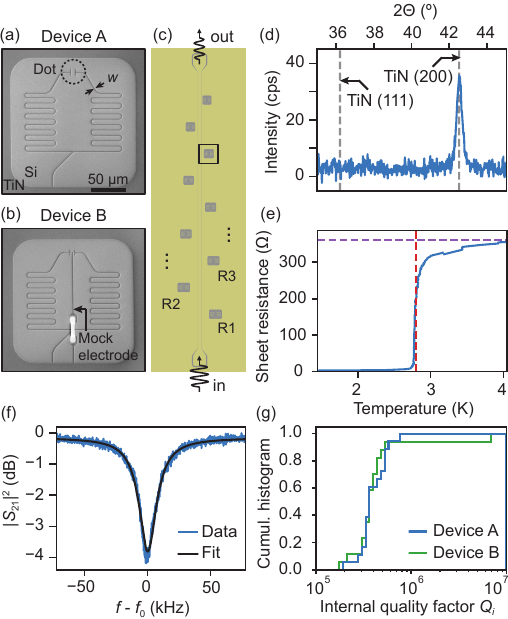}
    \caption{Experimental realization of SCRs (a) Scanning electron micrograph of a fabricated TiN (dark gray) resonator on Si substrate (light gray). The wire width $w = 1.6$~$\mu$m. (b) To test how control electrodes affect the resonance frequency and $Q_i$, we test an additional realization (Device B), which contains a mock electrode and air bridge along the resonator symmetry axis (c) Layout of a test chip containing nine resonators of varying meander length $\ell$. The resonators are coupled to a common microwave feedline to measure $S_{21}$. (d) X-ray diffraction spectrum for a 100~nm thick TiN film grown on Si (100) only shows evidence of the TiN (200) crystallographic orientation. The intensity was corrected for Si (100) background counts. (e) dc resistance measured across a test structure shows $T_c = 2.80\pm 0.05$~K (vertical dashed line). Using the measured sheet resistance $R_\square = 361 \pm 26$~$\Omega$ (horizontal dashed line) we find $L_\square = 178 \pm 13$~pH/$\square$. (f) Sample resonance for one of the resonators on device A with $f_0 = 5.025$~GHz. (g) Cumulative histogram of internal quality factors of 18 measured resonances for device A (blue, median $Q_i = 3.9 \times 10^5$) and device B (green, median $Q_i = 3.9 \times 10^5$). This shows $Q_i$ remains high even after adding a mock electrodes.}
    \label{fig:fig3}
\end{figure}

The physical implementation of the SCR circuit is shown in Fig.~\ref{fig:fig3}a and consists of two meandering wires of length $\ell$ galvanically connected to the ground plane via the tail. In order to maximize the coupling $g$, we aim to increase the inductance $L$ while decreasing the capacitance $C$, which results in an overall enhanced differential mode impedance $Z_{\mathrm{res}, d}$ as outlined in Eq.~\eqref{eq:diff_impedance}. 

To maximize $L$ we select titanium nitride (TiN) for the resonator material, because thin TiN films are known to have a high sheet inductance $L_{\square}$ with low loss at microwave frequencies \cite{Shearrow2018}. For ease of fabrication, we pick a wire width $w = 1.6$~$\mu$m which is compatible with optical lithography. This choice of $w$ simultaneously reduces the phase slip rate as a source of loss, because $w$ greatly exceeds the superconducting coherence length \cite{Mooij2015, Joshi2022}. We then adjust the length $\ell$ to obtain resonance frequencies in the range $3-6$~GHz. 

The resonator design in Figs.~\ref{fig:fig3}a,b does not have an explicit capacitor, and instead relies on the capacitance from the meandering wire to the far removed ground plane. In addition to reducing the resonator capacitance to 10~fF, the meandering reduces the footprint of our devices to less than $250\times 180$~$\mu$m$^2$. Finally, the length of the tail is designed to split the common and differential mode frequencies by less than 150 MHz. While in this work we short the inductive tail to the ground plane, in future experiments it can be connected to a dc electrode in order to bias the resonator center pin.

\begin{table*}[bthp]
\caption{Comparison of preferred TiN growth orientation on Si substrates for various deposition methods and deposition conditions. Si and TiN refer to the silicon substrate and TiN crystallographic orientations, $T$ is the deposition temperature, and method is the deposition method. ALD = atomic layer deposition, PECVD = plasma-enhanced chemical vapor deposition.}
\label{tab:summary_tin_on_si}
\begin{tabular}{@{}lcccc@{}}
\toprule
\multicolumn{1}{l}{Reference}                             & Si                     & TiN                                             & $T$ ($^\circ$C)      & Deposition method               \\ \toprule 
This work                                                 & (100)                  & (200)                                           & 270                  & ALD           \\ \midrule
Shearrow \emph{et al.} \cite{Shearrow2018}                & (111)                  & (200)                                           & 270                  & ALD          \\ \midrule
\multirow{2}{*}{Vissers \emph{et al.} \cite{Vissers2010}} & \multirow{2}{*}{(100)} & (111)                                           & 20                   & \multirow{2}{*}{Sputtering}     \\
                                                          &                        & (200)                                           & 500                  &                                 \\ \midrule
Vaz \emph{et al.} \cite{Vaz2002}                          & (100)                  & $\delta$-phase                                  & 300                  & Sputtering                      \\ \midrule
\multirow{2}{*}{Ohya \emph{et al.} \cite{Ohya2014}}       & \multirow{2}{*}{(001)} & (111) at low stress                             & \multirow{2}{*}{20}  & \multirow{2}{*}{Sputtering}     \\
                                                          &                        & (200) at high stress                            &                      &                                 \\ \midrule
Narayan \emph{et al.} \cite{Narayan1992}                  & (100)                  & (200)                                           & 600-700              & Laser physical vapor deposition \\ \midrule
\multirow{2}{*}{Oh \emph{et al.} \cite{Oh1993}}           & \multirow{2}{*}{(100)} & (200) for low plasma power                      & \multirow{2}{*}{400} & \multirow{2}{*}{PECVD}          \\
                                                          &                        & \multicolumn{1}{l}{(111) for high plasma power} &                      &                                 \\ \bottomrule
\end{tabular}
\end{table*}

We create two different device implementations: Device A (Fig.~\ref{fig:fig3}a) and Device B. Device B has the same basic design as Device A, but also includes a mock TiN electrode connected via an aluminum airbridge that crosses over the resonator near the tail (Fig.~\ref{fig:fig3}b). In future experiments that involve electrons on helium, this mock electrode will help to trap electrons in dot and tune the dot curvature $\omega_\mathrm{dot}$. The airbridge is deliberately placed along the resonator symmetry axis at an ac voltage node to minimize its effect on the resonance. In testing the microwave properties, we verify that the resonator characteristics are indeed unaltered.

We fabricate the resonators following the process developed in Ref.~\cite{Shearrow2018} using thin films of TiN deposited on Si (100) via plasma-enhanced atomic layer deposition at 270~$^{\circ}$C. We perform 150 cycles of atomic layer deposition (ALD), corresponding to an approximate film thickness of 12 nm. Further fabrication details can be found in Appendix~\ref{app:resonator_fab}. 

TiN can grow in the (200) and (111) crystallographic orientations, where the (200) orientation is associated with higher quality factors and lower microwave losses \cite{Vissers2010}. To verify the chemical composition of our devices, we perform X-ray diffraction (XRD) spectroscopy on both the TiN film and on the bare silicon substrate (Fig.~\ref{fig:fig3}d). The background-subtracted intensity displays a Gaussian diffraction peak at $2\Theta= 42.5^\circ$, corresponding to the (200) orientation of TiN \cite{Liu2024, Vissers2010, Shearrow2018, Jaim2015, Vaz2002, Oh1993}, while no notable signal from TiN (111) is observed. The (200) orientation is consistent with other works where TiN is grown at elevated temperatures ($>$200$^{\circ}$C) with differing deposition techniques (see Table~\ref{tab:summary_tin_on_si}) \cite{Chang2013, Liu2024, Vissers2010, Oh1993, Narayan1992}, suggesting that temperature-related film strain and surface energy, and not the Si substrate orientation, are the key determining factors for the preferred film orientation \cite{Vissers2010, Chang2013, Liu2024, Oyetoro2024}. 

The resulting sheet inductance $L_\square$ is found from a four point probe resistance measurement across a narrow TiN strip, which reveals a critical temperature of $T_{c}=2.80\pm 0.05$~K and sheet resistance of $R_\square = 361 \pm 26$~$\Omega/\square$ (Fig ~\ref{fig:fig3}e). From these measurements, we determine a sheet inductance $L_\square = \hbar R_\square / 1.76 \pi k_B T_c = 178\pm 13$~pH/$\square$. Compared to previous TiN films fabricated with ALD on Si (111) \cite{Shearrow2018}, our devices have a lower $T_{c}$ which indicates a higher degree of disorder and results in higher overall sheet inductance $L_\square$.

\subsection{Resonator microwave characterization}
To investigate the microwave properties of the resonators, test chips are fabricated with multiple resonators that are capacitively coupled to a single feed line (Fig.~\ref{fig:fig3}c), enabling measurement of all resonators in parallel. We design each chip to have 9 resonators with resonance frequencies between 3-6~GHz, tuned by varying the length $\ell$. In addition, we vary the distance to the feed line to control the coupling capacitance. 

After fabrication, we characterize the microwave properties of the SCRs at $T=10$~mK in a dilution refrigerator, well below the critical temperature ($T_c=2.8$~K). We probe the microwave transmission $S_{21}$ with an input power corresponding to an average intra-cavity photon number $\bar{n} \approx 10^0-10^2$. The transmission spectrum shows a total of 18 resonances, consistent with each of the 9 resonators having a common and differential mode. Near each resonance, the transmission amplitude $S_{21}$ follows \cite{Megrant2012}
\begin{equation} \label{complexS21}
S_{21} = \left( 1 + \frac{Q_i}{Q_c} e^{i\varphi} \frac{1}{1 + 2i Q_i (f - f_0) / f_0} \right)^{-1},
\end{equation}
where $Q_{i}$ is the internal quality factor, $Q_{c}$ is the coupling quality factor, $\varphi$ is a phase due to a feedline-resonator impedance mismatch, $f$ is the frequency, and $f_{0}$ is the resonator frequency. From this expression we extract $f_0$, $Q_c$ and $Q_i$ for each resonance (Fig.~\ref{fig:fig3}f), and plot the statistical distribution of $Q_i$ for device A and B in Fig.~\ref{fig:fig3}g. 

\begin{figure*}[htbp]
    \centering
    \includegraphics[width=1.5\columnwidth]{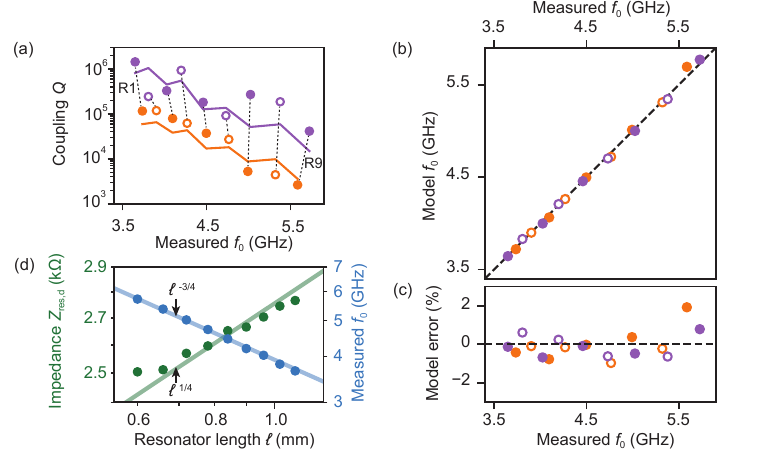}
    \caption{SCR model verification for device A resonances (a) Coupling quality factors for all 18 measured resonances (dots). Eq.~\eqref{eq:qc_from_circuit_model} (lines) predicts a higher $Q_c$ for the differential mode (purple) compared with the common mode (orange), which we use for classification. Closed (open) dots are used for odd (even) resonator numbers Rn throughout this figure. The overall decrease of $Q_c$ towards higher $f_0$ is by design and reflects the increasing coupling capacitance to the feedline. (b) Predicted common (orange) and differential (purple) mode resonances using the model described in the main text. The model inputs are the electrostatic simulations of the capacitance matrix and the measured TiN sheet inductance. A fit to the factor $\gamma$ results in $\gamma = 0.61 \pm 0.01$, where the error bar comes from fit uncertainty. (c) Difference between model $f_0$ and measured $f_0$. The lack of structure suggests that the model accurately reflects the fabricated resonators. (d) Log-log plot of differential mode frequencies (blue, right axis) and impedances (green, left axis) vs. resonator length $\ell$. Solid lines are guides to the eye from a model that predicts $f_0 \propto \ell^{-3/4}$ and $Z_{\mathrm{res},d} \propto \ell^{1/4}$.}
    \label{fig:fig4}
\end{figure*}

The internal quality factors of all resonators exceed $Q_{i} \sim 10^{5}$, and the median internal quality factor of both Device A and Device B is $Q_{i} = 3.9 \times 10^{5}$, corresponding to a median resonator internal loss rate $\kappa_i/2\pi = 11.7$~kHz. For more details on the power dependence of $Q_i$ see Appendix~\ref{app:Qi_power_dependence}. Since the $Q_i$ statistics for device A and B are similar, we conclude that the presence of the mock electrode and airbridge (Fig.~\ref{fig:fig3}c) does not limit $Q_i$ for both devices. Furthermore, it suggests that both devices have a common origin limiting $Q_i$, such as dielectric loss introduced during fabrication of both devices, or losses exacerbated by oxidation of TiN \cite{Logothetidis1999, Nahar2016}.

\subsection{Resonator model verification}

The impedance of the differential mode is a key resonator property that determines the electron-photon coupling strength, but is challenging to measure directly. Therefore, we resort to simulation of the resonator inductance and capacitance matrix, and map the experimental design to the equivalent circuit model of Fig.~\ref{fig:fig1}a. By comparing the simulated resonance frequencies with measured values, we quantify how accurately the circuit model reflects the experimental resonator design, and eventually, find the impedance of the differential mode.

Finding the differential mode impedance first requires identifying which of the 18 measured resonances are differential and common modes. To do this, we study the extracted values for $Q_c$, which show an alternating pattern of high and low $Q_c$ (Fig.~\ref{fig:fig4}a). This pattern can be explained through the SCR circuit model, which predicts \cite{Houck2008}
\begin{align}
	Q_c \approx \frac{1}{Z_{\mathrm{res},c,d} Z_0 C_c^2 \omega_0^2}, \label{eq:qc_from_circuit_model}
\end{align}
where $Z_{\mathrm{res},c,d}$ is the common or differential mode impedance, $Z_0$ is the microwave feedline impedance, and the effective coupling capacitance $C_c$ depends on the RF mode overlap between the resonator mode and feedline, which takes on the form $C_{c}=\frac{1}{2}(C_{c,a}\pm C_{c,b})$ for the common and differential modes, respectively. Here, $C_{c,a}$ and $C_{c,b}$ are the coupling capacitances of the feed line to each half of the SCR, see Fig.~\ref{fig:fig_s3}. Because the differential modes have much smaller $C_c$, and the mode splitting for all resonators was designed  $<$150 MHz, we can easily identify the differential modes by their relatively high $Q_c$ (Fig.~\ref{fig:fig4}a, purple dots). Having identified the differential modes, we find no statistical difference between common and differential mode $Q_i$. 

To quantitatively determine the extent to which the SCR circuit model describes the experimental resonators, we extract the capacitance matrix using finite element modelling software, and determine the inductances from the wire geometry and measured sheet inductance. Next, we diagonalize the circuit Hamiltonian and compare the predicted eigenfrequencies with the measured frequencies (see App.~\ref{app:q3d_modeling} and Table~\ref{tab:q3d_simulation} for details). We find that taking the raw values of the capacitance matrix yields inaccurate predicted eigenfrequencies, because the distributed capacitance of the meandering wire does not perfectly reflect the circuit model. To address this, we discount the simulated capacitances for all resonators by a single parameter $\gamma < 1$, which accounts for the distributed nature of the capacitance \cite{Kamenov2020, NiepcePRAppl2019, Shearrow2018}. By comparing measured and simulated eigenfrequencies, and treating $\gamma$ as a fit parameter we find $<2$\% error for all 18 resonators provided $\gamma = 0.61 \pm 0.01$ (Fig.~\ref{fig:fig4}b,c). The small error, combined with the lack of structure in the error confirms the SCR circuit model accurately represents the fabricated resonators. We note that the common-differential mode splitting changes sign at increased resonator lengths due to the decreasing ratio of the tail inductance to the total inductance (see Appendix \ref{app:inductive_tail} for details).

Finally, we extract the differential mode characteristic impedance $Z_{\mathrm{res},d} = L \omega_{0, d}$, which varies between 2.5 and 2.8~k$\Omega$ (Fig.~\ref{fig:fig4}d, green). Interestingly, the lower frequency (longer $\ell$) resonators have higher impedance, because the resonator inductance increases linearly with $\ell$, while the meandering causes the capacitance to scale with the resonator perimeter, and therefore $C \propto \sqrt{\ell}$ \cite{Kamenov2020, Shearrow2018}. As a consequence, the measured resonance frequencies are predicted to scale as $f_0 \propto \left(w / \ell^3 \right)^{1/4}$ and $Z_{\mathrm{res},d} = \left( \ell /w^3 \right)^{1/4}$, which is in excellent agreement with the experimental data. The scaling of the frequencies with $\ell$ falls between coplanar waveguide resonators ($f_0 \propto \ell^{-1}$) and lumped element resonators ($f_0 \propto \ell^{-1/2}$), reflecting the unique resonator design. Overall, the high impedance of all fabricated devices suggests a $\sqrt{2500/50} \approx 7\times$ boost (Eq.~\eqref{eq:coupling_strength}) in electron-resonator coupling compared to a standard 50~$\Omega$ device, making this device geometry highly desirable for applications that require strong coupling to single electrons.

\section{Discussion} \label{sec:discussion}

It is possible to increase the resonator impedance further by reducing the wire width $w$. From the scaling of $w$ and $\ell$ (Fig.~\ref{fig:fig4}d), we estimate that resonators with similar resonance frequency ($\omega_0' = \omega_0$), with half the length ($\ell' = \ell/2$) and an $8 \times$ reduced width ($w' = w/8$) give a 
\begin{equation}
    \frac{Z'}{Z} = \frac{\ell'}{\ell} \frac{w}{w'} \frac{\omega_0'}{\omega_0} = 4
\end{equation}
times higher impedance ($Z' \approx 10$~k$\Omega$) than those shown in Fig.~\ref{fig:fig4}d. The corresponding resonator capacitances are achieved by adjusting the wire width and length only, and therefore do not require adjustments to the ground plane separation. This would amount to a $14\times$ boost in coupling strength compared with a 50~$\Omega$ resonator, which comes at the moderate expense of having to use electron beam lithography for fabrication. For a similar dot design as in Ref.~\cite{Koolstra2019}, this would amount to $g/2\pi \approx 80$~MHz, which approaches the typical coupling strength of superconducting qubits. In addition, as these impedances approach or surpass the resistance quantum $R_Q \approx 6.45$~k$\Omega$, such high inductances may also be of use in protected superconducting qubits \cite{Hazard2019, Gyenis2021}.

Studying the interaction of electrons on helium with SCRs requires the addition of helium microchannels to shuttle electrons towards the dot, and electrodes that control the electron motional frequency via tuning the dot electrostatic potential. Such electrodes may cause a decrease in quality factor if they cannot be placed along a symmetry axis, especially for the high impedance SCR design \cite{HarveyCollard2020}, which can be mitigated by effective on-chip low-pass filters \cite{Zhang2024}. In addition, effective tuning of the dot curvature necessitates biasing the resonator center pin. To do this, the SCR tail can terminate in an electrode connected to a dc voltage source. This dc connection would act as a resistive component to the effective circuit that damps and totally suppresses the common mode. In this case, any asymmetry in the resonator mode profile also results in a reduction of $Q_i$ for the differential mode. We estimate that a 1\% asymmetry in the capacitances causes the differential mode $Q_i$ to drop below $10^5$. Experiments involving resonators with integrated microchannels, biasing electrodes, and coupling of electrons to the differential mode are currently underway.

Future experiments that address the spin state of electrons on helium require an in-plane magnetic field to define a spin quantization axis \cite{Schuster2010}. While the performance of TiN resonators in magnetic fields is not well studied, the relatively low $T_c$ of TiN suggests other nitrides such as niobium nitride (NbN) \cite{Anferov2020, Yu2021, Frasca2023} and niobium titanium nitride (NbTiN) are better suited to withstand magnetic fields \cite{Samkharadze2016}. Our SCR design can be easily adapted for NbN and NbTiN, and it is possible to compensate for a potentially lower $L_\square$ by decreasing the wire width $w$. Narrower wires are also less likely to trap magnetic vortices, which leads to reduced $Q_i$ \cite{Samkharadze2016}. 

\section{Conclusions} \label{sec:conclusion}
In conclusion, we have designed, analyzed, and experimentally demonstrated a high-impedance resonator design for coupling to the in-plane motion of electrons on helium. Our fabricated devices display low losses and excellent agreement with theoretical modeling. In addition, the small footprint comfortably allows us to fit several resonators on a single chip, which will prove useful for future experiments involving individual readout of several electrons on helium \cite{Beysengulov2024}. Crucially, our current device predicts a sevenfold increase in coupling strength compared to standard 50~$\Omega$ resonators, which may provide the enhancement needed to reach the strong coupling regime, a key milestone for both orbital and spin-based electron-on-helium quantum processors.  

\section*{Acknowledgements}
We would like to thank J. Theis for technical support. XRD measurements were carried out by ProtoXRD laboratories, and we are grateful for their help interpreting the results. This work made use of the Pritzker Nanofabrication Facility of the Institute for Molecular Engineering at the University of Chicago, which receives support from Soft and Hybrid Nanotechnology Experimental (SHyNE) Resource (NSF ECCS-2025633), a node of the National Science Foundation’s National Nanotechnology Coordinated Infrastructure.

\appendix
\titleformat{\section}[block]{\bfseries\normalsize\centering}{APPENDIX \thesection:}{1em}{\MakeUppercase}
\renewcommand{\thefigure}{\arabic{figure}}
\renewcommand\theequation{\Alph{section}\arabic{equation}}
\renewcommand{\thesection}{\Alph{section}}
\renewcommand{\thesubsection}{\arabic{subsection}}
\setcounter{equation}{0}

\section{Resonator Lagrangian with inductive tail} \label{app:inductive_tail}
\begin{figure}[htbp]
    \centering
    \includegraphics[width=\columnwidth]{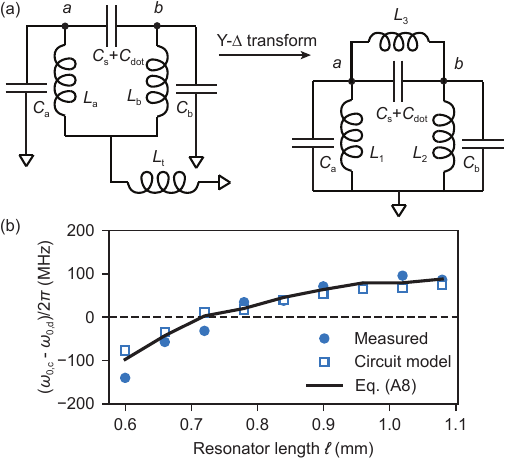}
    \caption{(a) Application of the Y-$\Delta$ transform for the circuit of Fig.~\ref{fig:fig1}, including the inductive tail. The transformed circuit is analytically tractable using the Lagrangian method, and includes an inductive coupling $L_3$. (b) Measured and simulated mode splitting for the 9 resonators. Circuit model refers to the simulation procedure outlined in App.~\ref{app:q3d_modeling}. Eq.~\eqref{eq:mode_splitting_symmetric_res} is an approximation of this procedure which assumes equal inductances and capacitances. In this equation, $C$ includes any coupling capacitance to the feedline, which can be found in Table~\ref{tab:q3d_simulation}.}
    \label{fig:fig_s2}
\end{figure}

A viable strategy to lift the degeneracy between the common and differential mode involves adding an inductor at the ground node between $L_a$ and $L_b$ as shown in Fig.~\ref{fig:fig_s2}a. This tail introduces a third circuit node, which complicates the circuit analysis. To simplify the circuit and reduce the number of nodes, we apply the Y-$\Delta$ transform to nodes $a$, $b$ and ground. The resulting circuit has two nodes and can now be analyzed efficiently. 

To find the eigenmodes of the transformed circuit, we find the Lagrangian, ignoring the offset field piercing the loop spanned by nodes $a,b$ and ground:
\begin{align}
    \mathcal{L}_\mathrm{res} &= \frac{C_a}{2}\dot{\phi}_a^2 + \frac{C_b}{2}\dot{\phi}_b^2 + \frac{C_x}{2} \left( \dot{\phi}_a - \dot{\phi}_b \right)^2 \notag \\
    &- \left[ \frac{\phi_a^2}{2 L_1} + \frac{\phi_b^2}{2 L_2} + \frac{(\phi_a - \phi_b)^2}{2L_3} \right] \label{eq:resoantor_lagrangian_with_tail}.
\end{align}
The inductors $L_{1, 2, 3}$ are linear combinations of $L_{a}$, $L_{b}$ and $L_{t}$. If $L_a = L_b = L$, we find
\begin{align}
    L_1 &= L_2 =  L + 2 L_t\\ 
    L_3 &= \frac{L (L + 2 L_t)}{L_t}.
\end{align}
The eigenfrequencies of this circuit are given by the matrix equation $\omega^2 \vec{\phi} =  \mathbf{C}^{-1} \mathbf{L}^{-1} \vec{\phi}$, where 
\begin{align}
    &\mathbf{C} = \begin{pmatrix} C_a + C_x & -C_x \\ -C_x & C_b + C_x  \end{pmatrix} \label{eq:lagrangian_w_tail_c}\\ 
    &\mathbf{L}^{-1} = \begin{pmatrix} \frac{1}{L_2} + \frac{1}{L_3} & - \frac{1}{L_3} \\ -\frac{1}{L_3} & \frac{1}{L_1} + \frac{1}{L_3} \end{pmatrix}. \label{eq:lagrangian_w_tail_l}
\end{align}
If the capacitors $C_a = C_b = C$ are also symmetric, this yields the two eigenfrequencies for the common ($\omega_{0, c}$) and differential ($\omega_{0,d}$) modes:
\begin{align}
    \omega_{0, c}^2 &= \frac{1}{(L + 2 L_t) C} \\
    \omega_{0, d}^2 &= \frac{1}{L(C + 2C_x)}.
\end{align}
These equations show that the common frequency tunes with $L_t$ while the differential frequency tunes with $C_x = C_s + C_\mathrm{dot}$. 

For small values of $L_t/L$ and $C_x/C$, the splitting between the common and differential modes is 
\begin{align}
    \omega_{0, c} - \omega_{0,d} = \frac{1}{\sqrt{LC}} \left( \frac{C_x}{C} - \frac{L_t}{L} \right), \label{eq:mode_splitting_symmetric_res}
\end{align}
and therefore the splitting can change sign depending on the shunt capacitance and tail inductance. For the resonators measured in this work, the splitting ranges from +100 MHz to -150 MHz (Fig.~\ref{fig:fig_s2}b), changing mainly due to increasing $L_t/L$ from R1 to R9 (see Table~\ref{tab:q3d_simulation}). 

\section{Theory of charge-photon coupling for the SCR design} \label{app:hamiltonian_formalism_coupling}
To study the interaction of trapped electrons with microwave photons in our SCR design and to show the promise of SCR for reaching strong coupling, we use the Hamiltonian formalism and calculate the eigenmodes of our coupled systems, which are ultimately shown in Fig.~\ref{fig:fig2}. In this section we derive and eventually diagonalize the total Hamiltonian
\begin{align}
    \mathcal{H}_\mathrm{tot} = \mathcal{H}_\mathrm{res} + \mathcal{H}_\mathrm{e} + \mathcal{H}_\mathrm{e-res},
\end{align}
which consists of the resonator $\mathcal{H}_\mathrm{res}$,  electrons $\mathcal{H}_\mathrm{e}$, and the coupling between the two $\mathcal{H}_\mathrm{e-res}$.

First, consider the resonator Hamiltonian. Using the Lagrangian from Eq.~\eqref{eq:resoantor_lagrangian_with_tail} as a starting point, we can define the node charges $q_n$ via the capacitance matrix,
\begin{equation}
    q_n = \sum_m \mathbf{C}_{nm} \dot{\phi}_m = \frac{\partial \mathcal{L}_\mathrm{res}}{\partial \dot{\phi}_n}.
\end{equation}
This allows us to construct the Hamiltonian in terms of node charges $q_n$ \cite{Vool2017}:
\begin{align}
    \mathcal{H}_\mathrm{res} = \frac{1}{D} &\left[ \frac{1}{2}(C_b + C_x) q_a^2 + \frac{1}{2}(C_a + C_x)q_b^2  + C_x q_a q_b \right] \notag \\ 
    & + \frac{\phi_a^2}{2L_1} + \frac{\phi_b^2}{2 L_2} + \frac{(\phi_a - \phi_b)^2}{2 L_3},
\end{align}
where $D = \mathrm{det}(\mathbf{C}) = C_a C_b + C_a C_x + C_b C_x$ is the determinant of the capacitance matrix.

Next, we derive the Hamiltonian for in-plane electron motion in the general case of $n$ trapped electrons, taking into account the electrostatic confinement potential and electron-electron interactions. For sufficiently thick helium films the electron-electron interactions are unscreened, i.e. pure Coulombic. This leads to the following Hamiltonian for the in-plane motion \cite{Yang2016}:
\begin{align}
    \mathcal{H}_\mathrm{e} = &\frac{1}{2m_e} \sum_i \left( p_{x, i}^2 + p_{y, i}^2 \right) - e \sum_i V_\mathrm{dc} (\mathbf{r}_i) \notag  \\ 
    + &\frac{1}{2} \frac{e^2}{4\pi \epsilon_0} \sum_{i}\sum_{j \neq i} \frac{1}{|\mathbf{r}_i - \mathbf{r}_j|}.
\end{align}
$p_{x, y, i}$ are the momenta of electron $i$ in the $x$ and $y$ directions, respectively, and $V_\mathrm{dc}$ is the electrostatic confinement potential. Expanding this Hamiltonian to second order around the equilibrium positions $\mathbf{r}_i = (x_i, y_i)$ results in
\begin{align}
    \mathcal{H}_\mathrm{e} = \frac{1}{2m_e} &\sum_i \left( p_{x, i}^2 + p_{y, i}^2 \right) \notag \\
    - \frac{e}{2} \sum_i &\left[ \frac{\partial^2 V_\mathrm{dc}}{\partial x^2} \delta x_i^2 + \frac{\partial^2 V_\mathrm{dc}}{\partial y^2} \delta y_i^2 + 2\frac{\partial^2 V_\mathrm{dc}}{\partial x \partial y} \delta x_i \delta y_i \right] \notag \\ 
    + \frac{1}{2} \sum_{i} \sum_{j \neq i} &\left[ k_{ij}^+ (\delta x_i - \delta x_j)^2 + k_{ij}^-(\delta y_i - \delta y_j)^2 \right. \notag \\ 
    &\left. + 2 l_{ij} (\delta y_i - \delta y_j)(\delta x_i - \delta x_j)    \right] \notag,
\end{align}
where $\delta x_i$ and $\delta y_i$ are small excursions around the equilibrium positions, and we have neglected energy offset terms that do not play a role in the dynamics of the electrons. Furthermore, the coefficients $k_{ij}^\pm$ and $l_{ij}$ characterize the Coulomb interaction strength, and are given by \cite{Yang2016} 
\begin{align}
    &k_{ij}^\pm = \frac{1}{4} \frac{e^2}{4\pi \epsilon_0} \frac{1 \pm 3 \cos \left( 2 \theta_{ij} \right) }{r_{ij}^3} \label{eq:eom_kij_plus}\\
    &l_{ij} = \frac{1}{4} \frac{e^2}{4\pi \epsilon_0} \frac{3 \sin \left( 2 \theta_{ij} \right)}{r_{ij}^3}, \label{eq:eom_lij}
\end{align}
where $r_{ij} = |\mathbf{r}_i - \mathbf{r}_j|$ is the pairwise distance and $\theta_{ij} = \arctan((y_i - y_j)/(x_i - x_j))$ is the angle between electron $i$ and $j$. Using the tools developed in Ref. \cite{Koolstra_Quantum_Electron_2024}, these geometric quantities can be evaluated for arbitrary electrostatic potentials.

Lastly, we derive the electron-resonator coupling term $\mathcal{H}_\mathrm{e-res}$, which captures the dipole coupling between resonator photons and electron motion. In contrast to previous work, which considered coupling of electrons to a single flux node \cite{Yang2016}, here we consider electrons coupling to two flux nodes (labeled $a$ and $b$), making our theoretical treatment more generally applicable to novel resonator designs. 

The $i$-th electron's energy at equilibrium position $\mathbf{r}_i$ due to a voltage on plate $a$ is given by $-e \alpha_\mathrm{res}^a (\mathbf{r}_i) \dot{\phi}_a$, where $\alpha_\mathrm{res}^a (\mathbf{r}_i)$ is the lever arm of plate $a$, which is determined from finite-element method (FEM) simulations by setting $\dot{\phi}_a = 1$~V while $\dot{\phi}_b = 0$. A similar argument holds for plate $b$, such that the total energy for electron $i$ sums to 
\begin{equation}
    \mathcal{H}_{\mathrm{e-res}, i} = -e \left( \alpha_\mathrm{res}^a (\mathbf{r}_i) \dot{\phi}_a + \alpha_\mathrm{res}^b (\mathbf{r}_i) \dot{\phi}_b \right)
\end{equation}
Summing over all electrons and linearizing around the equilibrium positions $\mathbf{r}_i$ yields 
\begin{align}
     \mathcal{H}_\mathrm{e-res} = -e \sum_{i} &\left[ \left( \frac{\partial \alpha_\mathrm{res}^a}{\partial x}(\mathbf{r}_i) \delta x_i + \frac{\partial \alpha_\mathrm{res}^a}{\partial y}(\mathbf{r}_i) \delta y_i \right) \dot{\phi}_a \right. \notag \\
    &\left. + \left( \frac{\partial \alpha_\mathrm{res}^b}{\partial x}(\mathbf{r}_i) \delta x_i + \frac{\partial \alpha_\mathrm{res}^b}{\partial y}(\mathbf{r}_i) \delta y_i \right) \dot{\phi}_b \right], \label{eq:el_res_ham_with_phidot}
\end{align}
where we have neglected the constant energy offset that does not affect the equations of motion. Because the current induced in each plate is proportional to the electron's velocity and electric fields $\partial \alpha_\mathrm{res}/ \partial x$ and $\partial \alpha_\mathrm{res} /\partial y$:
\begin{equation}
    \dot{q}_a = -\frac{d}{dt} \frac{\partial \mathcal{H}_\mathrm{e-res}}{\partial \dot{\phi}_a} = e \sum_i \mathbf{\nabla} \alpha_\mathrm{res}^a (\mathbf{r}_i) \cdot \dot{\mathbf{r}}_i,
\end{equation}
the coupling term of Eq.~\eqref{eq:el_res_ham_with_phidot} is consistent with the Shockley-Ramo theorem \cite{Shockley1938}. 

Because we have chosen node charge and node flux as our conjugate variables for $\mathcal{H}_\mathrm{res}$, we seek to express $\mathcal{H}_\mathrm{e-res}$ in terms of node charges. We use the relations from the definition of $\mathbf{C}^{-1}$ and substitute 
\begin{align}
    &\dot{\phi}_{a} = \frac{\partial \mathcal{H}_\mathrm{res}}{\partial q_a} = \frac{C_b + C_x}{D} q_{a} + \frac{C_x}{D} q_{b} \label{eq:hamil_phidot_b}\\
    &\dot{\phi}_b  = \frac{\partial \mathcal{H}_\mathrm{res}}{\partial q_b} = \frac{C_x}{D} q_{a} + \frac{C_a + C_x}{D} q_b \label{eq:hamil_phidot_c} 
\end{align}
into Eq.~\eqref{eq:el_res_ham_with_phidot}, which gives the desired form
\begin{align}
     \mathcal{H}_\mathrm{e-res} = -e \sum_{i} &\left[ \left( \beta_{x, i}^a \delta x_i + \beta_{y, i}^a \delta y_i \right) q_a \right. \notag \\
     &+ \left. \left( \beta_{x, i}^b \delta x_i + \beta_{y, i}^b \delta y_i \right) q_b \right]. \label{eq:H_el_res_simplified}
\end{align}
Note that we have defined the following shorthand notations
\begin{align}
    &\beta_{x, i}^a = \frac{C_b + C_x}{D} \frac{\partial \alpha_\mathrm{res}^a }{\partial x}(\mathbf{r}_i) + \frac{C_x}{D} \frac{\partial \alpha_\mathrm{res}^b}{\partial x}(\mathbf{r}_i) \\ 
    &\beta_{x, i}^b =  \frac{C_a + C_x}{D} \frac{\partial \alpha_\mathrm{res}^b}{\partial x}(\mathbf{r}_i) + \frac{C_x}{D} \frac{\partial \alpha_\mathrm{res}^a }{\partial x}(\mathbf{r}_i) \\ 
    &\beta_{y, i}^a = \frac{C_b + C_x}{D} \frac{\partial \alpha_\mathrm{res}^a }{\partial y}(\mathbf{r}_i) + \frac{C_x}{D} \frac{\partial \alpha_\mathrm{res}^b}{\partial y}(\mathbf{r}_i) \\ 
    &\beta_{y, i}^b = \frac{C_a + C_x}{D} \frac{\partial \alpha_\mathrm{res}^b}{\partial y}(\mathbf{r}_i) + \frac{C_x}{D} \frac{\partial \alpha_\mathrm{res}^a}{\partial y}(\mathbf{r}_i).
\end{align}

\subsection{Equations of motion for ensembles in two dimensions}
The joint dynamics of the electrons and resonator are governed by the Hamilton equations:
\begin{align}
    \dot{\phi}_n  &= \frac{\partial \mathcal{H}_\mathrm{tot}}{\partial q_n} & \dot{p}_{x, i} &= \frac{\partial \mathcal{H}_\mathrm{tot}}{\partial \delta x_i} & \dot{p}_{y, i} &= \frac{\partial \mathcal{H}_\mathrm{tot}}{\partial \delta y_i} \\
    \dot{q}_n  &= -\frac{\partial \mathcal{H}_\mathrm{tot}}{\partial \phi_n} & \dot{\delta x}_i  &= -\frac{\partial \mathcal{H}_\mathrm{tot}}{\partial p_{x, i}} & \dot{\delta y}_i  &= -\frac{\partial \mathcal{H}_\mathrm{tot}}{\partial p_{y, i}}  \label{eq:invL_eqs_with_H_el_res}
\end{align}
Here $n = a, b$ and $i$ denotes the electron index. The equations of motion for the resonator are 
\begin{align}
    &\dot{\phi}_a = \frac{C_b + C_x}{D} q_a + \frac{C_x}{D} q_b -  e \sum_{i} \left[ \beta_{x, i}^a \delta x_i + \beta_{y, i}^a \delta y_i \right] \label{eq:phi_dot_a}\\
    &\dot{\phi}_b  = \frac{C_x}{D} q_a + \frac{C_a + C_x}{D} q_b -  e \sum_i \left[ \beta_{x, i}^b \delta x_i + \beta_{y, i}^b \delta y_i   \right] \\ 
    &\dot{q}_a = -(L_1^{-1} + L_3^{-1}) \phi_a + L_3^{-1} \phi_b \\ 
    &\dot{q}_b = L_3^{-1} \phi_a - (L_2^{-1} + L_3^{-1}) \phi_b \label{eq:q_dot_b}
\end{align}
The equations of motion for $\dot{p}_{x,i}$ and $\dot{p}_{y,i}$ are
\begin{align}
    \dot{p}_{x_i} &= \left( -e \frac{\partial^2 V_\mathrm{dc}}{\partial x^2} + \sum_{j \neq i} k_{ij}^+ \right) \delta x_i - \sum_{j \neq i} k_{ij}^+ \delta x_j \notag  \\
    &+ \left( -e \frac{\partial^2 V_\mathrm{dc}}{\partial x \partial y} + \sum_{j \neq i} l_{ij} \right) \delta y_i - \sum_{j \neq i} l_{ij} \delta y_j \label{eq:p_dot_x_general} \\
    &- e \left( \beta_{x, i}^a q_a +\beta_{x, i}^b q_b \right) \notag  \\ 
    \dot{p}_{y_i} &= \left( -e \frac{\partial^2 V_\mathrm{dc}}{\partial y^2} + \sum_{j \neq i} k_{ij}^- \right) \delta y_i - \sum_{j \neq i} k_{ij}^- \delta y_j \notag  \\
    &+ \left( -e \frac{\partial^2 V_\mathrm{dc}}{\partial x \partial y} + \sum_{j \neq i} l_{ij} \right) \delta y_i - \sum_{j \neq i} l_{ij} \delta y_j  \\
    &- e \left( \beta_{y, i}^a q_a +\beta_{y, i} ^b q_b \right) \notag
\end{align}
Next, using Eqs.~\eqref{eq:invL_eqs_with_H_el_res}, we arrive at 
\begin{align}
    &\dot{\delta x}_{i} = -\frac{p_{x, i}}{m_e} \label{eq:x_dot_general}\\
    &\dot{\delta y}_{i} = -\frac{p_{y, i}}{m_e}
\end{align}

\subsection{Equations of motion for $n = 1$ in one dimension}
In the main text, we consider one electron trapped in a one-dimensional electrostatic potential of the form 
\begin{equation}
    -e V_\mathrm{dc} = \frac{1}{2} m_e \omega_\mathrm{dot}^2 x^2.
\end{equation}
The equations of motion follow from combining Eqs.~\eqref{eq:p_dot_x_general}-\eqref{eq:x_dot_general} and \eqref{eq:phi_dot_a}-\eqref{eq:q_dot_b}:
\begin{align}
    -\begin{pmatrix} \ddot{q}_a \\ \ddot{q}_b \\ \ddot{\delta x} \end{pmatrix} = \mathbf{\tilde{L}^{-1}} \mathbf{\tilde{C}^{-1}} \begin{pmatrix} q_a \\ q_b \\ \delta x \end{pmatrix},
\end{align}
where 
\begin{align}
     \mathbf{\tilde{L}^{-1}} &= \begin{pmatrix} L_1^{-1} + L_3^{-1} & -L_3^{-1} & 0 \\ -L_3^{-1} & L_2^{-1} + L_3^{-1} & 0 \\ 0 & 0 & m_e^{-1} \end{pmatrix} \\ 
     \mathbf{\tilde{C}^{-1}} &= \begin{pmatrix} (C_b + C_x)/D & C_x / D & -e \beta_x^a \\ C_x / D & (C_a + C_x)/D & -e \beta_x^b \\ -e \beta_x^a & -e \beta_x^b & m_e \omega_\mathrm{dot}^2 \end{pmatrix}.
\end{align}
Using the Ans\"atze $\delta x(t)  = \delta x e^{i \omega t}$ and $q_{a,b} (t) = q_{a,b} e^{i \omega t}$, we arrive at the eigenvalue equation $\omega^2 \mathbf{q} = \mathbf{\tilde{L}^{-1}} \mathbf{\tilde{C}^{-1}} \mathbf{q}$. The results of this diagonalization for various $\omega_\mathrm{dot}$ yield the $n=1$ curves shown in Fig.~\ref{fig:fig2}.

\subsection{Equations of motion for $n = 2$ in one dimension}
Two electrons in a one-dimensional harmonic potential naturally separate by a distance 
\begin{equation}
    d = \left( \frac{e^2}{2\pi \epsilon_0 m_e \omega_\mathrm{dot}^2} \right)^{\frac{1}{3}},
\end{equation}
due to the repulsive Coulomb interaction. The effects of this interaction on the in-plane electron motion are encoded in $k_{12}^+$ and $k_{21}^+$, which are given by
\begin{align}
    k_{12}^+ = k_{21}^+ = \frac{e^2}{4 \pi \epsilon_0}\frac{1}{d^3} = \frac{1}{2} m_e \omega_\mathrm{dot}^2,
\end{align}
since $\theta_{12} = 0$ in one dimension in Eq.~\eqref{eq:eom_kij_plus}. 

With these results, we may update $\mathbf{\tilde{L}^{-1}}$ and $\mathbf{\tilde{C}^{-1}}$:
\begin{align}
     \mathbf{\tilde{L}^{-1}} &= \begin{pmatrix} L_1^{-1} + L_3^{-1} & -L_3^{-1} & 0 & 0\\ -L_3^{-1} & L_2^{-1} + L_3^{-1} & 0 & 0 \\ 0 & 0 & m_e^{-1} & 0 \\ 0 & 0 & 0 & m_e^{-1} \end{pmatrix} \\ 
     \mathbf{\tilde{C}^{-1}} &= \begin{pmatrix} \frac{C_b + C_x}{D} & \frac{C_x}{D} & -e \beta_{x,1}^a & -e \beta_{x,2}^a \\ \frac{C_x}{D} & \frac{C_a + C_x}{D} & -e \beta_{x, 1}^b & -e \beta_{x, 2}^b \\ -e \beta_{x,1}^a & -e \beta_{x,1}^b & \frac{3}{2} m_e \omega_\mathrm{dot}^2 & -\frac{1}{2} m_e \omega_\mathrm{dot}^2 \\ -e \beta_{x,2}^a & -e \beta_{x,2}^b & -\frac{1}{2} m_e \omega_\mathrm{dot}^2 & \frac{3}{2} m_e \omega_\mathrm{dot}^2\end{pmatrix}.
\end{align}
Diagonalizing the matrix $\mathbf{\tilde{L}^{-1}} \mathbf{\tilde{C}^{-1}}$ for various $\omega_\mathrm{dot}$ yields the $n=2$ curve displayed in Fig.~\ref{fig:fig2}b.

\section{Coupling strength $g$ for the SCR design} \label{app:coupling_strength_derivation}
The electron coupling Hamiltonian for a single electron in one dimension is given by Eq.~\eqref{eq:el_res_ham_with_phidot}. Writing this coupling term in terms of $\dot{\phi}_c$ and $\dot{\phi}_d$, we find
\begin{equation}
    \mathcal{H}_{\mathrm{e-res}} = -\frac{e \, \delta x}{\sqrt{2}} \left[ \frac{\partial \alpha_\mathrm{res}^a}{\partial x} \left( \dot{\phi}_c + \dot{\phi}_d \right) + \frac{\partial \alpha_\mathrm{res}^b}{\partial x} \left( \dot{\phi}_c - \dot{\phi}_d \right) \right]
    \label{eq:el-res-coupling-commdiff}
\end{equation}
For an electron centered in the dot $\frac{\partial \alpha_\mathrm{res}^a}{\partial x} = - \frac{\partial \alpha_\mathrm{res}^b}{\partial x} = \mathcal{E}_x$, such that 
\begin{equation}
    \mathcal{H}_{\mathrm{e-res}} = - \sqrt{2} e \mathcal{E}_x \delta x \dot{\phi}_d.
    \label{eq:el-res-coupling-simplified}
\end{equation}
To relate this to Eq.~\eqref{eq:coupling_strength}, we use the expressions for the vacuum fluctuations of the electron motion $\delta x = \sqrt{\hbar / 2m_e \omega_e} \hat{x}$ and the resonator $\dot{\phi}_d = \omega_0 \sqrt{\hbar Z_{\mathrm{res},d}/2} \hat{V}$, where $\hat{x} = b + b^\dagger$ and $\hat{V} = a + a^\dagger$ are quantum operators. This yields 
\begin{equation}
    \mathcal{H}_{\mathrm{e-res}} = -  \frac{\sqrt{2}}{2} e \mathcal{E}_x \omega_0 \sqrt{\frac{Z_{\mathrm{res},d}}{m_e \omega_e}} \hat{x} \hat{V} = \sqrt{2} g  \hat{x} \hat{V}.
\end{equation}
Therefore, the coupling strength of SCR resonators is $\sqrt{2}$ larger than the simple $LC$-resonator.

We note that Eq.~\eqref{eq:el-res-coupling-simplified} only includes coupling to $\dot{\phi}_d$ in the case of symmetric capacitances $C_a = C_b = C$. For $C_a \neq C_b$, $\mathcal{H}_{\mathrm{e-res}}$ also includes a coupling to the common mode. This term is proportional to $\dot{\phi}_c$. For the asymmetries in this work, we estimate that the common mode coupling can become an appreciable fraction of the differential mode coupling. In future devices the common mode coupling can be suppressed through symmetric feedline coupling. Finally, any single-electron decoherence resulting from this coupling can be suppressed by increasing the common-differential mode splitting with a larger tail inductance $L_t$.

\section{Resonator fabrication} \label{app:resonator_fab}
TiN films are deposited on undoped high resistivity Si (100). Before TiN film deposition, the silicon wafers undergo a standard solvent clean followed by an oxide removal with hydrofluoric acid. Immediately after oxide removal, wafers are placed in a Ultratech/Cambridge Fiji G2 for Plasma-Enhanced atomic layer deposition of TiN. All devices characterized in this manuscript underwent 150 cycles of deposition, corresponding to a film thickness of approximately 12 nm.

After ALD, the resonator geometry is defined using a combination of optical lithography and etching. We perform optical lithography with a Heidelberg MLA150 Direct Writer to define the TiN etch mask. After lithography, we etch the TiN using a chlorine-based Plasma-Therm inductively coupled plasma etch recipe. The etch recipe combines Cl$_2$:BCl$_3$:Ar with flow rates 30:30:10~sccm, and uses 400~W ICP power and 50~W bias power at a 5~mTorr chamber pressure. The resist mask remaining after TiN etching  was removed  in AZ NMP rinse at 80~$^\circ$C, followed by rinsing with IPA.

To fabricate the mock TiN electrode and airbridges, we perform two additional rounds of lithography using the same optical process described above. The base resonator geometry and TiN portion of the mock electrode are defined in the same lithography and etch step. The fabrication process for the airbridges is adapted from Ref.~\cite{Chen2014}. With another round of lithography, we first form the scaffold to support Al airbridges with a 3~$\mu$m thick positive photoresist. The resist is then re-flowed at 140~$^\circ$C for better mechanical stability. Next, 300~nm thick aluminum is deposited at high~vacuum using a Plassys MEB550S Electron Beam Evaporator to form the metal layer of the airbridge. Before the aluminum deposition, in-situ ion milling is performed to remove the native oxide on the TiN surface for better ohmic contact between aluminum and TiN. Through a second round of lithography, the bridge pattern is defined with the same thick photoresist and the aluminum not used for the airbridge is removed by wet etching with Transene Aluminum Etchant Type A. Before aluminum wet etching, the resist is baked at 115~$^\circ$C to improve adhesion to the aluminum. After wet etching, the resist is removed by a combination of dry and wet resist strip processes. During the dry strip process, most of the cross-linked resist formed from the ion milling step prior to the aluminum deposition is removed by oxygen plasma. The sample is then soaked in AZ NMP rinse to remove all remaining resist, followed by rinsing with IPA.

Up to this point, both Device A and Device B undergo this same fabrication process, although only Device B has airbridges. Next, samples are diced and cleaned, where chips with airbridges are rinsed in IPA instead of DI water and blow dried to prevent collapse of the suspended structures from the surface tension of the water. Finally, the samples are mounted to a PCB, and wirebonded before installation onto the mixing chamber of the dilution refrigerator. 

\section{Experimental setup} \label{app:exp_setup}

\begin{figure}[htbp]
    \centering
    \includegraphics[width=\columnwidth]{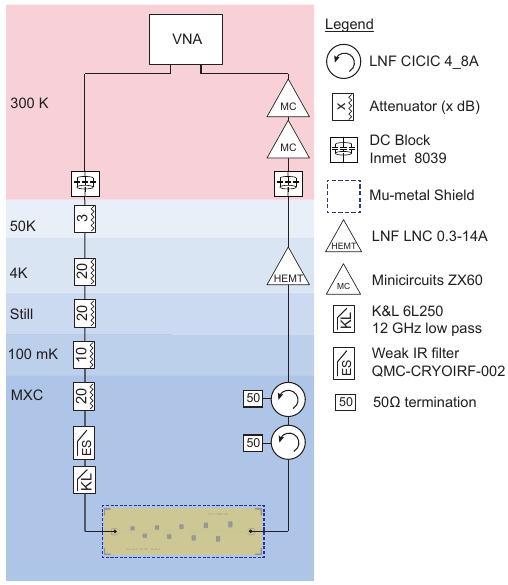}
    \caption{Experimental setup. Shaded areas represent different temperature stages of the dilution refrigerator. The microwave devices measured in this work are mounted to the mixing chamber plate (MXC), which operates at a temperature of 10~mK.}
    \label{fig:fig_s1}
\end{figure}

All measurements were performed in a Bluefors LD400 dilution refrigerator at a base temperature of 10~mK. Test chips were mounted and wirebonded to a Rogers interposer and placed in an RF enclosure from Lotus Communications before mounting to the cryostat mixing chamber baseplate. Microwave transmission measurements were performed using a Keysight E5071C network analyzer. Microwave signals traveling from room temperature are attenuated by -73~dBm using cryogenic attenuators. Additional low-pass and infrared filters at the bottom of the microwave chain suppress any thermal photons. Next, the transmission passes through an LNF-CICIC4\_8A double junction circulator and is amplified by a LNF-LNC0.3\_14A high-electron mobility transistor (HEMT) amplifier. The full microwave measurement chain is shown in Fig.~\ref{fig:fig_s1}. 

\section{Probe power dependence of $Q_i$} \label{app:Qi_power_dependence}
\begin{figure}[htbp]
    \centering
    \includegraphics[width=\columnwidth]{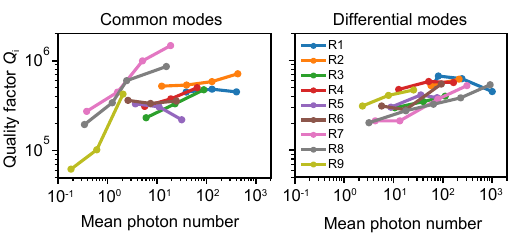}
    \caption{Extracted quality factors vs. resonator photon occupation $\bar{n}$. The quality factors are color-coded by resonator (R1-R9) and divided by mode type (common vs. differential). See Appendix.~\ref{app:Qi_power_dependence} for more details.}
    \label{fig:qi_vs_nbar}
\end{figure}

We show the power dependence of all device A resonances in Fig.~\ref{fig:qi_vs_nbar}. All $Q_i$ are extracted below the bifurcation power. The mean photon number $\bar{n}$ was inferred from \cite{Frasca2023}
\begin{equation}
    \bar{n} = \frac{P_\mathrm{in} Q_c}{\hbar \omega_0 ^2 } \left( \frac{Q_i}{Q_i + Q_c} \right)^2,
\end{equation}
which takes into account the power at the sample, $P_\mathrm{in}$, which is estimated from the VNA output power and the line attenuation of approx. 75~dB. Apart from the common mode of R9, all resonances only show a limited power dependence and have quality factors above $10^5$. 

\section{Modeling of resonator properties} \label{app:q3d_modeling}
\begin{figure}[htbp]
    \centering
    \includegraphics[width=\columnwidth]{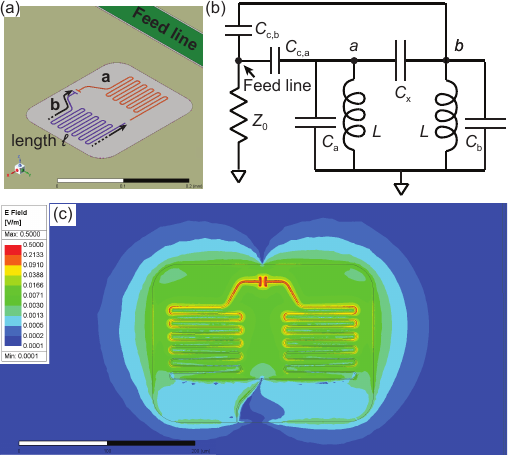}
    \caption{(a) Finite Element Model with circuit nodes. To accurately compare to the approximate circuit model, we simulate the capacitance matrix without the explicit connection between nodes $a$ and $b$, and without connections between nodes $a$ and $b$ to ground. (b) Approximate circuit used to simulate the resonator design in this work, which includes coupling to a feed line with characteristic impedance $Z_0 \approx 50$~$\Omega$. (c) Simulated electric field profile (magnitude, log scale) for the differential mode of R1.}
    \label{fig:fig_s3}
\end{figure}

\begin{table*}[]
\caption{Simulated resonator properties for design A. $\ell$ is the length of the meandering wire, $C_{a,b}$ are the capacitances of each meander to ground, $C_x$ is the total shunt capacitance, $C_{c,a}$ and $C_{c,b}$ are the coupling capacitances to the feed line, $L_t$ is the tail inductance, and $L$ is the inductance of the meandering wire.}
\label{tab:q3d_simulation}
\begin{tabular}{@{}ccccccccc@{}}
\toprule
 & $\ell$ (mm) & $C_a$ (fF) & $C_b$ (fF) & $C_x$ (fF) & $C_{c,a}$ (fF) & $C_{c,b}$ (fF) & $L$ (nH) & $L_t$ (nH) \\\midrule
R1 & 1.08        & 21.6      & 21.8      & 1.70       & 0.6           & 0.3           & 117.1   & 6.5  \\
R2 & 1.02        & 20.8       & 20.9       & 1.68    & 0.5           & 0.3           & 110.1    & 6.5  \\
R3 & 0.96        & 19.9       & 20.3       & 1.68      & 0.7           & 0.4           & 103.8    & 6.5  \\
R4 & 0.90        & 19.3       & 19.5       & 1.62      & 0.6           & 0.4           & 97.4    & 6.5  \\
R5 & 0.84        & 18.0       & 18.5       & 1.55      & 1.0           & 0.5           & 91.0    & 6.5  \\
R6 & 0.78        & 17.5       & 17.9       & 1.49      & 0.9           & 0.4           & 84.7    & 6.5  \\
R7 & 0.72        & 16.2       & 16.9       & 1.47       & 1.3           & 0.6           & 77.7    & 6.5  \\
R8 & 0.66        & 15.5       & 16.3       & 1.40       & 1.2           & 0.5           & 71.3    & 6.5  \\
R9 & 0.60        & 13.8       & 15.0       & 1.33      & 2.0           & 0.7           & 64.9    & 6.5  \\\bottomrule
\end{tabular}
\end{table*}

To accurately compare our designed resonators to the circuit model of Fig.~\ref{fig:fig1}b, we first simulate the capacitance matrix in Ansys Electronics Desktop (see Fig.~\ref{fig:fig_s3}a), which outputs values for $C_{a, b}$, $C_x$ and coupling to the feedlines $C_{c, a}$ and $C_{c, b}$. Next, we calculate the inductance, taking into account the measured TiN sheet inductance $L_\square$:
\begin{align}
    L = L_\square \frac{\ell}{w}, \label{eq:total_calc_inductance}
\end{align}
where $\ell$ and $w$ are the nanowire length and width, respectively. We ignore the geometric inductance and mutual inductance because these effects are small compared with the kinetic inductance \cite{QiMingChen2023}. Table~\ref{tab:q3d_simulation} shows simulated capacitances and calculated inductances. 

Before feeding the tabulated values into the resonator model, we first discount all capacitances by a factor of $\gamma$ to account for the distributed nature of the mode profile along the meandering wire. To justify this factor $\gamma$, consider the case of a simple $\lambda/2$ resonator. In this case we can map the $\lambda/2$ resonator to a lumped element resonator via their analytical expressions for both resonance frequencies:
\begin{align}
    f_0 = \frac{1}{2\pi \sqrt{L_\mathrm{tot} C_\mathrm{tot}}} = \frac{1}{4 \sqrt{(L_\ell \ell) (C_\ell \ell)}}, 
\end{align}
where $L_\mathrm{tot}$, $C_\mathrm{tot}$ are the inductance and capacitance participating in the resonance, and $L_\ell$, $C_\ell$ are the effective inductance and capacitance per unit length. Note that $\ell$ is half the total resonator length (see Fig.~\ref{fig:fig_s3}a),  $L_\ell$ is found from Eq.~\eqref{eq:total_calc_inductance} and $C_\ell \ell$ is the simulated capacitance $C_\mathrm{sim}$ found from Ansys Electronics Desktop. Therefore, in the case of a straight $\lambda/2$ resonator we have 
\begin{align}
    C_\mathrm{tot} = \left(  \frac{2}{\pi} \right)^2 C_\ell \ell = \gamma \, C_\mathrm{sim},
\end{align}
and thus $\gamma = (2/\pi)^2 \approx 0.40$. On the other hand, in the case of a pure lumped element resonator, no correction needs to be applied to map the resonator onto the circuit model of Fig.~\ref{fig:fig1}, and therefore $\gamma = 1$. Therefore, $\gamma$ is practically bounded in the range $(2/\pi)^2 \leq \gamma \leq 1$. By treating $\gamma$ as a fit parameter we find the degree to which the mode approximates the lumped circuit model. 

Finally, we must also take into account the feedline capacitances. We approximate the effect of the feedline capacitances by adding them to $C_a$ and $C_b$: $C_a \mapsto C_a + C_{c,a}$ and $C_b \mapsto C_b + C_{c,b}$.

We can now diagonalize the matrix equation $\omega^2 \vec{\phi} =  \mathbf{C}^{-1} \mathbf{L}^{-1} \vec{\phi}$, where $\mathbf{C}^{-1}$ and $\mathbf{L}^{-1}$ are given by Eqs.~\eqref{eq:lagrangian_w_tail_c}-\eqref{eq:lagrangian_w_tail_l}. The resulting eigenfrequencies take into account $C_a, C_b$ and $C_s$ and contain the effects of feedline capacitances as well as tail inductance.  We classify the eigenfrequencies by common or differential mode through inspection of the eigenvectors, and finally, we compare the measured frequencies to our prediction. The results are plotted in Fig.~\ref{fig:fig4}. 

An alternative way to model the resonances is by FEM simulations in the microwave domain. We have found this method generally more cumbersome than the capacitance matrix method, but it can provide additional insights. After applying a sheet reactance boundary condition corresponding to 180~pH/sq, Fig.~\ref{fig:fig_s3}c shows the resulting eigenmode field profile for the differential mode of R1. This mode is characterized by a non-vanishing electric field between the meandering wires. In addition, the eigenmode frequency is 3.62 GHz, consistent with the measured value and the capacitance matrix model.

\bibliography{thebibliography}

\end{document}